\documentclass[journal,onecolumn,12pt]{IEEEtran}

\usepackage{amsmath,amssymb,amsfonts,mathrsfs}
\usepackage{algorithmic}
\usepackage{graphicx}
\usepackage{textcomp}
\usepackage{xcolor}
\usepackage{amsthm}
\usepackage{amsmath}
\newtheorem{theorem}{Theorem}

\newtheorem{corollary}[theorem]{Corollary}

\usepackage{amsmath,blkarray,booktabs,bigstrut}
\usepackage{tikz}
\usepackage{color}
\usepackage{color, colortbl}
\definecolor{Gray}{gray}{0.9}
\usepackage{caption}
\theoremstyle{definition}
\newtheorem{definition}[theorem]{Definition}
\newtheorem{remark}[theorem]{Remark}
\newtheorem{example}[theorem]{Example}
\usepackage{blkarray}
\usepackage{nomencl}
\makenomenclature
\renewcommand{\nomname}

\ifCLASSINFOpdf
\else
\fi
\begin{document}
%
\title{Single Server Private Information Retrieval Protocols With Codes Over Rings}
%
%
%

\author{\c{S}eyma~Bodur \dag \thanks{\dag Corresponding author: seyma.bodur@uva.es},
        Edgar~Mart\'inez-Moro,
        and~Diego~Ruano\\[1em]
        IMUVA-Mathematics Research Institute,\\
        Universidad de Valladolid, Spain
\thanks{This work was supported in part by Grant TED2021-130358B-I00 funded by MCIN/AEI/10.13039/501100011033 and by the ``European Union NextGenerationEU/PRTR'', and by Grant CONTPR-2019-385 funded by Universidad de Valladolid and Banco Santander.}
}

%
%

\markboth{}{}

%



\maketitle

\begin{abstract}
 A Private Information Retrieval  (PIR) protocol based on coding theory for a single server is proposed. It provides computational security against linear algebra attacks, addressing the main drawback of previous PIR proposals based on coding theory. The approach involves two types of codes each one over a different ring, an inner non-free linear code that will be used as a distinguisher of some elements added to the query matrix, and an outer code that will be used for generating the query matrix. Moreover, it only uses modular arithmetic at the server level and the recovering stage if the base ring chosen for the inner code is $\mathbb Z_m$.
\end{abstract}


%
\IEEEpeerreviewmaketitle

\section{Introduction}

Private Information Retrieval (PIR) protocols were first introduced in \cite{Sudan} and they allow users to retrieve data from a database without revealing which part of the database they are interested in. They can be used to provide privacy to users of generative artificial intelligence tools based on large language models as ChatGPT \cite{chatgpt}. The analysis of these protocols involves two main approaches: considering that the information is stored in one or several servers. For the several servers approach, one may consider information-theoretic privacy and they are secure against unlimited computation sources, they were studied in \cite{Dvir,2002Breaking,sun2017capacity, sun2018capacity,banawan2018capacity,freij2017private}. However, for the single server case, it is not possible to achieve information-theoretic privacy except for the naive solution of downloading the entire database \cite{Sudan}. A PIR protocol is considered computationally secure if the database administrator cannot determine which file is accessed by the user with a reasonable amount of computational resources, even if they know the query. This is usually known as the \textit{Distinguishability Problem}. This paper concentrates on the single-server approach and we only consider single-server PIR protocols from now on. 

Several PIR protocols are available in the literature, however, all of them have important drawbacks. The protocols in \cite{kushilevitz1997replication, lipmaa2017simpler} are based on some cryptographic features, such as the integer factorization problem, and they will become insecure with quantum computers. Furthermore, in \cite{sion2007computational} it was proved that a protocol based on number theoretic issues is less efficient than downloading the entire database.
In \cite{yi2012single,kiayias2015optimal,melchor2016xpir,gentry2019compressible}, the authors proposed PIR protocols using fully homomorphic encryption. Aiming at single-server PIR protocols, all those PIR protocols depend on the database size in terms of communication cost. When the size becomes larger, the communication will not be efficient, although the query size of \cite{melchor2016xpir} was reduced in \cite{angel2018pir}. The first PIR protocol based on coding theory was presented by Holzbaur, Hollanti, and Wachter-Zeh in \cite{9174138} and it is efficient, but Bordage and  Lavauzelle presented in  \cite{Bordage} a polynomial-time attack to this protocol that success with high probability. 

This paper provides a single server PIR protocol based on coding theory over rings which is resistant to the attack in \cite{Bordage} based on detecting a rank difference in the query matrix by deleting rows. Our protocol can be understood as a modification of the protocol in \cite{9174138} and it will be based on codes over finite rings. We take advantage of the fact that some of the involved codes are non-free as ring modules, which makes linear algebra attacks non-feasible.  We will require two finite rings $R\subset\mathcal R$,  such that $\mathcal R$ can be seen as a $R$-submodule. We will define in $\mathcal R$ an  $R$-linear code called inner code and we will also establish a $\mathcal R$-linear code as the outer code.  All the transmitted information and the computations needed will be made with $R$ as the alphabet whereas the information of the codes will be used in the query generation and response processing, both by the user. We propose to consider $R=\mathbb Z_m$, the set of integers modulo a composite number $m$, and $\mathcal R= \mathbb Z_m[x]/\langle x^n-1\rangle$, with $\mathrm{gcd}(n,m)=1$. As inner code, we will use a $\mathbb Z_m$-cyclic code that can be seen as a Chinese remainder construction of its prime components (depending on the prime factorization of $m$) and the outer code will be a matrix-product code with constituents in $\mathcal R$ that turns out to be a quasi-cyclic code when it is seen over $\mathbb Z_m$.

 The hardness of a brute force attack will rely on the capacity of the server to guess the inner code, thus it depends on the knowledge of $n$ as well, which is not a public parameter, and on the fact that there is a big enough number of possible cyclotomic cosets that define different cyclic codes in $\mathcal R= \mathbb Z_m[x]/\langle x^n-1\rangle$. The protocol we propose can resist the rank difference attack known for single-server PIR protocols based on codes whereas our rate information is worse than the one in \cite{9174138}. To enhance security against the rank difference attack, we somewhat compromise the rate in the PIR protocol. Furthermore, as stated before, all the computations will be made as modular operations (modulo $m$) and therefore they are less computationally intense than other protocols that require large field extensions. Moreover, the server only has to perform a modular multiplication of matrices. 
 
The structure of the paper is as follows: In Section~\ref{sec:1} we briefly introduce the protocol \cite{9174138}. In Section~\ref{sec: 3}, we present our computational PIR protocol with codes over rings. Finally, in Section~\ref{sec:4}  we discuss the PIR rate and security analysis of our protocol.

\section{Single Server PIR Protocols}\label{sec:1}
 We will focus on the case that a single server is responsible for storing all files and has complete knowledge of them, therefore the task of the  PIR protocol is to ensure that the identity of the data requested by a user remains private. In a single-server PIR protocol, the process comprises three stages: query generation, file request, and response processing. In the query generation stage, a user initiates by determining a file they plan to retrieve from the database, called the \textit{desired file}. Subsequently, the user creates a query to obtain the desired file and transmits it to the server as a file request. Using the provided query, the server processes it according to the protocol and sends a response back to the user, thereby enabling the recovery of the desired file after processing it by the user. The difficulty for the server to know the desired file relies on the hardness of the so-called \textit{Distinguishability problem}, which consists on, given the query, determining the desired file.

\subsection{HHWZ PIR protocol}\label{sec:HHWZ}
In this section, we will briefly review the HHWZ PIR protocol in \cite{9174138}.  Let $q$ be a prime power, an $[n,k]_q$ linear code $C$ is a $k$-dimensional subspace of $\mathbb{F}_q^ n$ where $\mathbb{F}_q$ is the finite field with $q$ elements. We will denote by $G$  a generator matrix of $C$. An information set of the linear code $C$ is a set of  $k$ coordinates such that the elements in $C$ restricted to that set are all the elements in  $\mathbb F_q^k$. More precisely, $I \subseteq [n]:=\{1, \ldots , n\}$, with $| I | =k$, is an information set of $C$ if the $k \times k $ submatrix of $G$, whose columns are indexed by $I$, is a full rank matrix.

In the single server HHWZ PIR protocol \cite{9174138}, the setup is the following.  There are $t$ files stored on the server,  $m_i$ for $i \in \{1,\ldots,t \}$. For simplicity, we consider here that the files are a vector of length $L$ with entries in $\mathbb{F}_q$. 

Let $s>0$ be an integer and let $\{ \mathbf b_1,\mathbf b_2,\ldots,\mathbf b_s \}$ be a basis of $\mathbb{F}_{q^s}$ as a vector space over $\mathbb{F}_q$. Let us  consider $V=\langle \mathbf b_1,\ldots, \mathbf b_v\rangle_{\mathbb{F}_q}$ and $V^c=\langle \mathbf b_{v+1},\ldots,\mathbf b_s\rangle_{\mathbb{F}_q}$, two linear subspaces. The 
user selects a random $[n,k]_{q^s}$ code $C \subseteq \mathbb{F}^n_{q^s} $, and let $\mathbf W$ be a matrix of size $ t \times n$ over $\mathbb{F}_{q^s}$ whose rows are selected uniformly at random from codewords in $C$. For an information set  $I$, of the code $C$, the user selects a matrix  $\mathbf E$ of size $t \times n$ and entries in the subspace  $V$,  such that a column in $\mathbf E$ is the all-zero vector if its index is in $I$. Finally, the user chooses a $t \times n$ matrix $\mathbf U$ whose elements $\mathbf U_{i,j}$  are in $V^c$ and  $\mathbf U_{i,j}$ is a non-zero element when  $i$ points to the desired file and  $j\in I$; and the entries $\mathbf U_{i,j}$ are zero elsewhere.  
The query matrix is given by $\mathbf{Q} = \mathbf W + \mathbf  E + \mathbf  U$, which the user sends to the server. The server computes $\mathbf r=\sum_{i=1}^{t} m_i \cdot \mathbf q_i$ where $\mathbf q_i$ represents rows of the query matrix and then sends the response $\mathbf r$ to the user.

Based on the previous fixed information set $I$, the user can eliminate the codewords of $C$ from the response matrix:
\begin{equation*}
    \mathbf r-\mathbf r_{I}G_{I}^{-1}G=\sum_{j\in I} m_i\cdot \mathbf i_j,
\end{equation*} where $\mathbf{i}_j$ is the $j$-th coordinate vector in $\mathbb F_{q^s}^n$.
Consider the projection over the subspace $V$ given by
\begin{eqnarray*}
    \textrm{proj}: \mathbb{F}_{q^s} &\longrightarrow& \mathbb{F}_{q^s}\\
     \sum_{i=1}^{s} x_i \cdot \mathbf b_i  &\longmapsto&  \sum_{i=1}^{v} x_i \cdot \mathbf b_i.
\end{eqnarray*}
Then, the desired file can be recovered by applying the projection of $\mathbb{F}_{q^s}$ over $V^c$ for the $j^{th}$ coordinate. 
 \begin{equation*}
     \textrm{proj}(\sum_{i=1}^{t} m_i \cdot\mathbf  e_i^j) =  m_d \cdot \mathbf e_j^d,
 \end{equation*}
where $\mathbf e^d$ is the $d^{th}$ row of the matrix $\mathbf E$. As the user knows $\mathbf e_j^d$, the desired file can be recovered.

It was shown in \cite{Bordage} an attack based on removing a row from the query matrix and checking whether the resulting matrix has a lower rank than the original matrix. In this way, the server may determine the desired file with a high probability. Our protocol avoids this attack.  

\section{Single Server PIR protocol Over Rings} \label{sec: 3}

In this section, we describe our PIR protocol. We will consider two rings $R\subset\mathcal R$ such that $\mathcal R$ is an $R$-module, and two types of codes, namely a code over $\mathcal R$ that will play a similar role as the code $C$ in  HHWZ PIR protocol and a $R$-code in $\mathcal R$ as an $R$-submodule. We will call the code over $\mathcal R$ the \emph{outer code}
and the $R$-subcode of $\mathcal R$ the \emph{inner code}.
We will follow  the following notations in this paper:\\

\nomname{\(C_\mathtt{IN}\)}{\quad The inner code, it is an $R$-subcode of $\mathcal R$.}
\nomname{\(C_\mathtt{OUT}\)}{\; The outer code, it is an $\mathcal R$-code}
\nomname{\(G_\mathtt{IN}\)}{\quad Matrix with entries in $\mathcal R$  whose rows generate the inner code.}
\nomname{\(G_\mathtt{OUT}\)}{\; Matrix with entries in $\mathcal R$  whose rows generate the outer code.}
\nomname{\(L\)}{\qquad Number of rows of the files in the database.}
\nomname{\(n\)}{\qquad Length of the inner code, ie. the number of generators of $\mathcal R$ as an $R$-module.}
\nomname{\(r\)}{\qquad Number of columns of the files in the database.}
\nomname{\(s\)}{\qquad Number of generators of outer code,}
\nomname{\(t\)}{\qquad Number of files in the database.}

\printnomenclature
\

From now on we will fix the following notation:  a  vector  (respectively a matrix) with entries in $\mathcal R$ will be denoted with round brackets,  $(\, )$.
Note that, since $\mathcal R$ is an $R$-submodule if we fix an $R$-generating set of  $\mathcal R$ any element $r\in \mathcal R$ can be expressed as a  row vector in $R^n$, where $n$ is the number of generators of $\mathcal R$ as $R$-module, we will denote that expansion with square brackets,  $[\, ]$. In the same fashion, this can be done componentwise to vectors and matrices with entries in  $\mathcal R$ and we will also use square brackets for that row-wise expansion of the entries of matrices and vectors.

 More concretely, throughout this paper, we will fix the ring $\mathcal R$, i.e. the alphabet for the query generation process, to be  $\mathcal{R}=R [x]/\langle x^{n}-1\rangle$. The alphabet for the projection like will be $R=\mathbb{Z}_m$, the set of integers modulo $m$ for a fixed $m\in \mathbb Z$. Moreover, we will take \(C_\mathtt{IN}\) to be an ideal in $\mathcal R$, that is, a  $\mathbb{Z}_m$ cyclic code of length $n$. As usual, one  can describe cyclic codes in $\mathcal R$ using the Chinese remainder theorem (CRT), that is, if $m= \Pi_{i=1}^{\ell} p_i^{e_i}$ is the prime decomposition of $m$ then
\begin{eqnarray*}
    \Phi : \frac{\mathbb{Z}_m[x]}{\langle x^{n}-1\rangle}&\longrightarrow&\bigoplus_{i=1}^{\ell}\frac{\mathbb{Z}_{p_i^{e_i}}[x]}{\langle x^{n}-1\rangle} \\
    \sum_{j=0}^{n-1} c_j x^j&\longmapsto&(\sum_{j=0}^{n-1}c_j(\bmod{p_1^{e_1}})x^j,\ldots, \sum_{j=0}^{n-1}c_j(\bmod{p_{\ell}^{e_{\ell}}}) x^j ) ,
\end{eqnarray*}
is a module isomorphism. Moreover, if we let $C_1,C_2,\ldots, C_s$ be cyclic codes in  $\mathbb{Z}_{p_i^{e_i}}[x]/ \langle x^{n}-1\rangle$, for $i \in \{1,\ldots,s\}$, then the set $\mathrm{CRT}(C_1,\ldots,C_s)= \{\Phi^{-1}(v_1,v_2,\ldots,v_s) \mid v_i \in C_i  \}$ is a $\mathbb Z_m$-cyclic code of length $n$. That is, it corresponds to an ideal in $\mathcal{R}=\mathbb{Z}_{m}[x]/ \langle x^{n}-1\rangle$. From \cite[Remark~2]{Bhowmick2020},  we have that
$\mathrm{CRT}(C_1,\ldots,C_s)$  is a non-free $\mathbb Z_m$-module  if and only if there is at least a pair $i,j\in\{1,2,\ldots, s\}$, with $i \neq j$,  such that $\mathrm{rank}_{p_i^{e_i}}(C_i) \neq \mathrm{rank}_{p_j^{e_j}}(C_j)$, where $\mathrm{rank}_{p_i^{e_i}}$ denotes the rank as $\mathbb Z_{p_i^{e_i}}$-module.

The inner code in our system $C_\mathtt{IN}$ will be a non-free cyclic code over $\mathbb{Z}_m$ of length $n$ that will be used for encoding.
Besides the code being non-free there is a matrix that can play the role of a  parity check matrix (see \cite{ChainRings}) that can be used as a distinguisher of whether an element in $\mathcal R$ is in $C_\mathtt{IN}$   or in $\mathcal R\setminus C_\mathtt{IN}$. In the case that $C_\mathtt{IN}$  is an LCD code there is a projection in the alphabet as in the HHWZ PIR protocol \cite{9174138}. Therefore, in our example, $C_\mathtt{IN}$ will be then a non-free LCD code using the results in \cite{Bhowmick2020}.

As an outer code $C_\mathtt{OUT}$, we will consider a linear code of length $s$ over $\mathcal R$. Note that, since $\mathcal R$ is chosen to be a cyclic ambient space, the set $[C_\mathtt{OUT}]=\{ [\mathbf v]\mid (\mathbf v)\in C_\mathtt{OUT}\}$ is a quasi-cyclic code over $\mathbb Z_m$ of length $ns$. A linear code $C$ of length $n {s}$ over the ring $R$ is said to be a quasi-cyclic code of index $s$ if it is invariant under the cyclic shift of codewords by ${s}$ positions and ${s}$ is the smallest number with this property.  We will define the outer code through its expansion, $[C_\mathtt{OUT}]$ over $\mathbb Z_m$, using a matrix-product code.

\begin{definition} [\cite{blackmore2001matrix,asch}]\label{def:matrixproduct}
    Let $C_1,\ldots,C_s  \subset \mathbb{Z}_m^n$ be linear codes of length $n$ and rank $r_i$, and let $M$ be an $s \times \ell$ matrix with entries in $\mathbb{Z}_m$. The matrix-product code $C=[C_1,..,C_s]M$ is defined as the set of all products $[c_1,..,c_s]M$ where $c_i \in C_i$ for $i \in \{1,..,s \} $. The code $C$ has  length $n\ell$, and if the matrix $M$ has rank $s$, then $C$ has rank $\mathrm{r_1}+\cdots+\mathrm{r_s}$.
\end{definition}

 Let $\Tilde{C_1},\Tilde{C_2},\ldots,\Tilde{C_s} \subset \mathcal{R}$, where $\Tilde{C_i}$ is a cyclic code over $\mathbb Z_m$ for $i \in \{1,\ldots,s\}$. Let $[C_\mathtt{OUT}]$ be the matrix-product code $[C_\mathtt{OUT}] = [\Tilde{C_1},\Tilde{C_2},\ldots,\Tilde{C_s}] M$, where $M$ is an $s \times s$   matrix over $\mathbb{Z}_m$. Hence, the outer code $C_{\mathtt{OUT}}$ is a $\mathcal{R}$ submodule in $\mathcal{R}^s$ that is a $s$-generator quasi-cyclic code. We will denote by $G_{\mathtt{OUT}}$ the generator matrix over $\mathcal{R}$.

 For technical reasons in the recovery stage and the security analysis, we will require that our codes fulfill the following technical conditions:
  \begin{itemize}
     \item The constituent codes are nested:  $\Tilde{C_1} \supseteq \Tilde{C_2}\supseteq\cdots\supseteq\Tilde{C_s}$.
     \item For $i \in \{1,\ldots,s\}$, $\Tilde{C_i} \cap C_\mathtt{IN} \neq \{ 0 \}$, and $\Tilde{C_i} \cap (C_\mathtt{IN}^{\bot} \setminus C_\mathtt{IN}) \neq \{ 0 \}$. 
     \item $e_i>1$, for all $i=1,\ldots, \ell$.
     \item The projections of the codes $\Tilde{C_1},\Tilde{C_2},\ldots,\Tilde{C_s}$ over $\mathbb Z_{p_i^{e_i}}$  are non-Hensel lifts (see Remark~\ref{rem:nh} for a definition), for $i=1,\ldots, \ell$.
 \end{itemize}

\begin{remark} Note that other possibilities for inner and outer codes may be chosen. That is, $C_\mathtt{IN}\subset \mathcal R$,  and $C_\mathtt{OUT}$ being a code with alphabet $\mathcal R$, can be achieved in many other situations. For example,  the broader class of codes over affine algebras with a finite commutative chain coefficient ring in \cite{affine}, that includes uni-(multivariate) codes (cyclic, negacyclic, constacyclic, polycyclic, abelian \ldots), quasi-cyclic, quasi-abelian, and many other can be used in this protocol.
\end{remark}

\subsection{Data Storage Process}
We assume that the data alphabet in the server is $\mathbb Z_{m^\prime}\subset \mathbb Z_m$ where $m^\prime= \Pi_{i=1}^{\ell} p_i$. The server will  contain $t$ files stored as an $ L \times r$ matrix of elements in  $\mathbb{Z}_{m^\prime}$ denoted as $\mathbf{DB^i} = [D^{i}_{j\ell}]$  where $i \in \{1,\ldots,t\}$, $j \in \{1,\ldots,L\}$, $\ell \in \{1,\ldots,r\}$ and all files are stored concatenated in the matrix   $\mathbf{DB}= [\mathbf{DB^1}\mathbin\Vert \mathbf{DB^2}\mathbin\Vert\cdots \mathbin\Vert \mathbf{DB^t}]$, where  the number of columns $r$ of each file is lower than or equal to $s$, the number of generators of $C_\mathtt{OUT}$, i.e. $r\leq s$. Throughout the paper, 
  $\mathbf{DB^d}$   will denote the desired file that the user wants to retrieve.

\tikzset{every picture/.style={line width=0.75pt}} 
\begin{center}

\begin{tikzpicture}[x=0.75pt,y=0.75pt,yscale=-1,xscale=1]

\draw  [fill={rgb, 255:red, 155; green, 155; blue, 155 }  ,fill opacity=0.55 ] (193.6,112.2) -- (235.6,112.2) -- (235.6,152.2) -- (193.6,152.2) -- cycle ;
\draw  [fill={rgb, 255:red, 155; green, 155; blue, 155 }  ,fill opacity=0.55 ] (235.6,112.2) -- (277.6,112.2) -- (277.6,152.2) -- (235.6,152.2) -- cycle ;
\draw  [fill={rgb, 255:red, 155; green, 155; blue, 155 }  ,fill opacity=0.55 ] (277.6,112.2) -- (319.6,112.2) -- (319.6,152.2) -- (277.6,152.2) -- cycle ;
\draw  [fill={rgb, 255:red, 155; green, 155; blue, 155 }  ,fill opacity=0.55 ] (319.6,112.2) -- (361.6,112.2) -- (361.6,152.2) -- (319.6,152.2) -- cycle ;
\draw  [fill={rgb, 255:red, 155; green, 155; blue, 155 }  ,fill opacity=0.55 ] (361.6,112.2) -- (403.6,112.2) -- (403.6,152.2) -- (361.6,152.2) -- cycle ;
\draw  [fill={rgb, 255:red, 155; green, 155; blue, 155 }  ,fill opacity=0.55 ] (403.6,112.2) -- (445.6,112.2) -- (445.6,152.2) -- (403.6,152.2) -- cycle ;

\draw (200,123.0) node [anchor=north west][inner sep=0.75pt]  [font=\small]  {$\mathbf{DB^{1}}$};
\draw (241.4,123.0) node [anchor=north west][inner sep=0.75pt]  [font=\small]  {$\mathbf{DB^{2}}$};
\draw (282.8,123.0) node [anchor=north west][inner sep=0.75pt]  [font=\small]  {$\mathbf{DB^{3}}$};
\draw (330,127.4) node [anchor=north west][inner sep=0.75pt]  [font=\small]  {$\cdots $};
\draw (373.2,127.4) node [anchor=north west][inner sep=0.75pt]  [font=\small]  {$\cdots $};
\draw (412.4,123.0) node [anchor=north west][inner sep=0.75pt]  [font=\small]  {$\mathbf{DB^{t}}$};
\draw (230.13,93.91) node [anchor=north west][inner sep=0.75pt]  [font=\Huge,color={rgb, 255:red, 128; green, 128; blue, 128 }  ,opacity=1 ,rotate=-89.31]  {$\{$};
\draw (209,87.8) node [anchor=north west][inner sep=0.75pt]  [font=\footnotesize]  {$r$};
\draw (440.93,93.31) node [anchor=north west][inner sep=0.75pt]  [font=\Huge,color={rgb, 255:red, 128; green, 128; blue, 128 }  ,opacity=1 ,rotate=-89.31]  {$\{$};
\draw (418.8,88.2) node [anchor=north west][inner sep=0.75pt]  [font=\footnotesize]  {$r$};
\draw (462.82,150.34) node [anchor=north west][inner sep=0.75pt]  [font=\Huge,color={rgb, 255:red, 128; green, 128; blue, 128 }  ,opacity=1 ,rotate=-181.34]  {$\{$};
\draw (461.6,123.6) node [anchor=north west][inner sep=0.75pt]  [font=\small]  {$L$};
\end{tikzpicture}   
\end{center}

\subsection{Query Generation}\label{sec:PIR}

First, the user sets up the codes.  The inner code  $C_\mathtt{IN}\subset \mathcal R=\mathbb{Z}_m[x]/\langle x^n -1 \rangle$ is a   cyclic $\mathbb{Z}_m$-linear code of length $n$. The user also chooses $C_\mathtt{OUT}$ as an $s$-generator quasi-cyclic code in  $\mathcal{R}^s$ arising from a matrix product code $[\Tilde{C_1},\Tilde{C_2},\ldots,\Tilde{C_s}] M$, such that $\Tilde{C_1},\Tilde{C_2},\ldots,\Tilde{C_s} \subset \mathcal{R}$, where each $\Tilde{C_i}$ is a cyclic code over $\mathbb Z_m$ for $i \in \{1,\ldots,s\}$ with the technical conditions stated above. We will denote by  $G_\mathtt{OUT}$ a generator matrix of $C_\mathtt{OUT}$  as an  $\mathcal{R}$-linear code.

The query generation begins with the user randomly selecting $s \cdot t \cdot r $ elements $a^i_{kj}$ in the  $m^\prime\mathcal{R}$. The user arranges these elements as $t$ matrices of size $s\times r$  $\mathbf{a}^i$, where the rows consists of $s$-uples in $\mathcal R^s$ 

$$\mathbf{a}^i=
\left(
\begin{array}{cccc}
 \mathrm{a^i_{11}} & \mathrm{a^i_{12}} &\ldots &\mathrm{a^i_{1s}}      \\
 \mathrm{a^i_{21}} & \mathrm{a^i_{22}} &\ldots &\mathrm{a^i_{2s}}      \\
\vdots &\vdots &&\vdots \\
 \mathrm{a^i_{r1}} & \mathrm{a^i_{r2}} &\ldots &\mathrm{a^i_{rs}}    
\end{array}
\right),$$
where $i\in\{1,\ldots,t\}$, $k\in\{1,\ldots,r\}$, and $j\in \{1,\ldots,s\}$.  Then the user encodes $\mathbf{a}^i$  as $$\mathbf{w}^i= \mathbf{a}^i\cdot G_\mathtt{OUT}, $$   i.e. the rows in $\mathbf{w}^i$ are codewords of $C_\mathtt{OUT}$.  Note that, in this encoding, all the multiplications are carried out in the ring $\mathcal R$.

Now the user selects $t$ matrices    $\mathbf{e}^i$ of size $r\times s$, for $i\in \{1,\ldots,t\}$, where each entry  $e^i_{kj} \in \mathrm{nf} (C_\mathtt{IN})$, for each $i\in \{1,\ldots,t\}$, $k\in \{1,\ldots,r\}$ and $j\in \{1,\ldots,s\}$, where $\mathrm{nf}(C)$ denotes the  non-free part of the code $C$.   
Moreover, the user  randomly selects a column position  $\gamma\in\{1,..,s-r+1\}$ and constructs $t$ matrices 
  $\mathbf{u}^i$ of size $r\times s$, for $i\in \{1,\ldots,t\}$,  with all zero entries but 
 \begin{equation}
      \mathrm{u^{d}_{1+\lambda,\gamma+\lambda}}\in \mathrm{nf}(\Tilde{C}_{s} \cap (C_\mathtt{IN}^{\bot} \setminus C_\mathtt{IN})), \; \text{for all} \; \lambda = 0,\ldots,r-1.
\end{equation} 
We recall that $d$ is the index of the file the user wants to retrieve from the server. Thus, $\mathbf{u}^i$ has zeros in all positions except in those entries belonging to the desired file. 
Let $\boldsymbol{\delta}^i=(\mathbf{w^i+e^i+u^i})$,  for $i\in\{1,\ldots,t\}$, and    $\boldsymbol{\Delta}$ and $\mathbf A$ the matrices with entries in $m^\prime\mathcal R$ with $r\cdot t$ rows given by
    
    \begin{equation}\label{delta}
    \boldsymbol{\Delta}=
    \left(
    \begin{array}{c}
\boldsymbol{\delta}^1 \\
\boldsymbol{\delta}^2\\
\vdots \\
\boldsymbol{\delta}^d\\
\vdots \\
\boldsymbol{\delta}^{t}
\end{array}
\right)
=
 \left(
    \begin{array}{c}
\mathbf{w^1+e^1+u^1} \\
\mathbf{w^2+e^2+u^2}\\
\vdots \\
\mathbf{w^{d}+e^{d}+u^{d}}\\
\vdots \\
\mathbf{w^{t}+e^{t}+u^{t}}
\end{array}
\right)
=
\left(
    \begin{array}{c}
\mathbf{w^1+e^1} \\
\mathbf{w^2+e^2}\\
\vdots \\
\mathbf{w^{d}+e^{d}+u^{d}}\\
\vdots \\
\mathbf{w^{t}+e^{t}}
\end{array}
\right), \qquad 
\mathbf{A}=
    \left(
    \begin{array}{c}
\mathbf{a}^1 \\
\mathbf{a}^2\\
\vdots \\
\mathbf{a}^d\\
\vdots \\
\mathbf{a}^{t}
\end{array}
\right).
 \end{equation}

  The user generates the query matrix $\mathbf{Q}$ with entries in $\mathbb Z_m$ by expanding in the ring $\mathbb Z_m$  the concatenation $(\mathbf{A}\mathbin\Vert \boldsymbol{\Delta})$, that is    ${\mathbf{Q}}=[\mathbf{A}\mathbin\Vert \boldsymbol{\Delta}]$. For a fixed  $i\in \{1,\ldots,t\}$, we will denote by $[\mathbf{q^i}]$, the submatrix of the query matrix whose rows are given by the rows $(i-1)r+1, \ldots ,ir$ in $\mathbf{Q}$, that is $[\mathbf{q^i}]=[\mathbf{a}^i\mathbin\Vert \boldsymbol{\delta}^i]$.

\subsection{Response}
Note that in the database, for a fixed $i\in \{1,\ldots,t\}$, the  $i^{th}$ file,  is an $r\times L$ matrix with entries in $\mathbb Z_m$ that we denote by $\mathbf{DB^i}$. Then the response is just the matrix multiplication over $\mathbb Z_m$ given by
\begin{equation}\label{eq:respZ}
\mathbf{R}=\mathbf{DB}\cdot \mathbf{Q},\quad \hbox{where } \mathbf{DB}=(\mathbf{DB^1} \mathbin\Vert \mathbf{DB^2} \mathbin\Vert \cdots \mathbin\Vert\mathbf{DB^t}).
\end{equation}

Note that, since $\mathbb Z_m\subset \mathcal R$, the response matrix  $\mathbf{R}$ that the server computes is just the $\mathbb Z_m$ expansion of the matrix
\begin{equation}\label{eq:respR}
   \sum_{i=1}^{t} \mathbf{DB^i}\cdot (\mathbf{a}^i\mathbin\Vert\boldsymbol{\delta}^i) =\sum_{i=1}^{t} (\mathbf{DB^i}\cdot \mathbf{a}^i\mathbin\Vert \mathbf{DB^i}\cdot (\mathbf{w}^i+\mathbf{e}^i+\mathbf{u}^i))=(\mathbf{R_1}\mathbin\Vert\mathbf{R_2}),
\end{equation}
which is a matrix with entries in $\mathcal R$, but the server does not know such decomposition since the ring $\mathcal R$ is kept secret.

\subsection{Recovering Stage}
We will use a recovering method that resembles the technique used in \cite{9174138} but without information sets.  The user can get the matrix $(\mathbf{R_1}\mathbin\Vert\mathbf{R_2})$ in Equation~(\ref{eq:respR})  from the matrix $\mathbf{R}$ in Equation~(\ref{eq:respZ}) since the user knows the ring $\mathcal R$ and henceforth $n$. Thus, the user can compute
\begin{equation}
    \mathbf{R_2}-(\mathbf{R_1}\cdot G_\mathtt{OUT}) = \sum_{i=1}^{t} (\mathbf{DB^i}\cdot \mathbf{w}^i+ \mathbf{DB^i}\cdot (\mathbf{e}^i+\mathbf{u}^i)- \mathbf{DB^i}\cdot \mathbf{w}^i)
    = \sum_{i=1}^{t} (\mathbf{DB^i}\cdot \mathbf{e}^i) + (\mathbf{DB^i}\cdot \mathbf{u^i}).
\end{equation}
Let us denote  by $\Gamma_s(C)=[C,\ldots,C]\mathrm{Id}_{s}$ the matrix product code of a cyclic code $C \subset \mathcal{R}$, where $\mathrm{Id}_s$ is the $s\times s$ the identity matrix and let us denote by $H_{\Gamma_{r}(C^{\bot}_{\mathrm{IN}})}$ a parity check matrix of the  code $\Gamma(C_{\mathrm{IN}})$ over $\mathbb Z_m$.

Then the user computes  
\begin{eqnarray*}
[\mathbf{R_2}-(\mathbf{R_1}\cdot G_\mathtt{OUT})]H^{\top}_{\Gamma_{r}(C^{\bot}_{\mathrm{IN}})}&=&\sum_{i=1}^{t} [\mathbf{DB^i}\cdot \mathbf{e}^i+\mathbf{DB^i}\cdot \mathbf{u}^i]H^{\top}_{\Gamma_{r}(C^{\bot}_{\mathrm{IN}})} \\
&=&\sum_{i=1}^{t}\underbrace{[\mathbf{DB^i}\cdot\mathbf{e}^iH^{\top}_{\Gamma_{r}(C^{\bot}_{\mathrm{IN}})}]}_{\quad=\;\mathbf{0}\;\hbox{\footnotesize  since}\;\mathbf e^i_{kj} \in C_\mathtt{IN}}+\underbrace{[\mathbf{DB^i}\cdot \mathbf{u}^iH^{\top}_{\Gamma_{r}(C^{\bot}_{\mathrm{IN}})}]}_{\quad=\; \mathbf{0},\;\hbox{\footnotesize when}\;t \neq d }\\
&=& [\mathbf{DB^d}\cdot \mathbf{u}^d H^{\top}_{\Gamma_{r}(C^{\bot}_{\mathrm{IN}})}]=\mathbf M.
\end{eqnarray*}
Note that the inner code provides us with a projection-like map in $\mathcal R$   (see Section~\ref{sec:HHWZ}).
Now, in order to retrieve the desired file the user must solve the following system of equations in $\mathbb Z_m$.
\begin{equation}
   \mathbf M= \begin{bmatrix} \mathbf{m_1} & & \\
 &  \ddots &\\
 & & \mathbf{m_r} 
\end{bmatrix}=\begin{bmatrix}
 \mathrm{x^{d}_{11}}&\cdots&  \mathrm{x^{d}_{1r}} \\
\vdots && \vdots \\
  \mathrm{x^{d}_{L1}}& \cdots & \mathrm{x^{d}_{Lr} }
\end{bmatrix} \cdot \begin{bmatrix}
 \mathrm{u^{d}_{1,\gamma}}H_{\mathrm{IN}}^{\top}  \\
&\mathrm{u^{d}_{2,\gamma+1}}H_{\mathrm{IN}}^{\top}  \\
 & & \ddots \\
 & & & \mathrm{u^{d}_{r,\gamma+r-1}}H_{\mathrm{IN}}^{\top} 
\end{bmatrix}
\end{equation}

We have that this system of equations over $\mathbb Z_m$ can be seen as several systems of linear equations, each one can be  expressed as
\begin{center}
$\mathbf{m_i}=$$\begin{bmatrix} 
\mathrm{x^{d}_{1i}} \\
\vdots \\
 \mathrm{x^{d}_{Li}}
\end{bmatrix}$$ \mathrm{u^{d}_{i,\gamma}}H_{\mathrm{IN}}^{\top} $, for $i \in \{1,\cdots,r\}$.
\end{center}
Note that $\mathbf{m_i}$  is a codeword in the code $\mathrm{u^{d}_{i,\gamma}}(C^{\bot}_{\mathrm{IN}})$, hence, once we fix a generator matrix for the code $C^{\bot}_{\mathrm{IN}}$ given by $H_{\mathrm{IN}}$, there is a unique expression for $\mathbf{m_i}$ in terms of the rows of the matrix $\mathrm{u^{d}_{i,\gamma}}H_{\mathrm{IN}}$ since the database elements are in $\mathbb Z_{m^\prime}$, which is the desired solution of the system.  For constructing the matrix $H_{\mathrm{IN}}$  we use the standard-like generator matrix for codes over that type of rings \cite{ChainRings} to build it via the inverse of the CRT. 

\begin{center}
\begin{figure}
   \centering
    \includegraphics[width=0.6\linewidth]{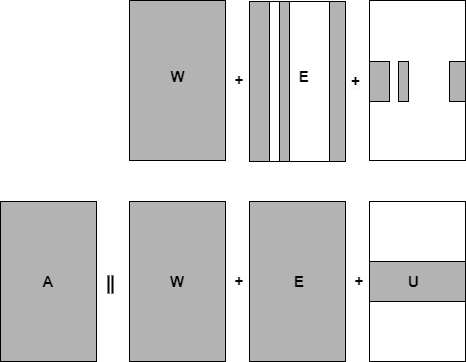}
    \caption{Query matrix in the original HHWZ PIR protocol vs. the modified one. White places means entries equal to $0$} \label{fig}
\end{figure}
\end{center}

\begin{remark}[{\bf This protocol over a field}]
Note that considering $C_\mathtt{IN}$ as a $\mathbb F_q$-code in $\mathbb F_q[x]/\langle f(x)\rangle$, with $f$ a primitive polynomial (i.e. an $\mathbb F_q$ subspace of $\mathbb F_q^{\mathrm{deg}(f)}$), and $C_\mathtt{OUT}$ as a code with alphabet $\mathbb F_{q^s}=\mathbb F_q[x]/\langle f(x)\rangle$, both codes operate within finite fields.  However, the arithmetic ceases to be modular if $q$ is not a prime. Note that this is nothing else than choosing the generators of $C_\mathtt{IN}$ as the basis of $V$ in HHWZ PIR protocol in Section~\ref{sec:HHWZ}. All the processes will follow as above and this will provide a PIR protocol with a database whose elements belong to the field $\mathbb F_q$. In this case, the query matrix will be larger than the one in HHWZ PIR protocol but it will not be safe against the rank difference attack in \cite{Bordage} as we will show in Section~\ref{sec:Security}. The primary distinction between the two protocols lies in the utilization of a generating set $A$, which eliminates the necessity for information sets. Consequently, the information contained in $\mathbf u$ can be distributed across the entire query matrix $\mathbf{Q}$, as shown in Figure~\ref{fig}.
\end{remark}

\begin{example}[Toy example]  For simplicity we will take $m'=m=3\cdot 5$, thus  $R=\mathbb Z_{15}$. Note that the value $m$ does not fulfill the technical conditions for security and that we are considering that $m'=m$, which is not allowed in our protocol. The aim is to illustrate the protocol with a toy example and simple computations.
  Consider the 3-file in database $\mathbf{DB}=[\mathbf{DB}^1, \mathbf{DB}^2, \mathbf{DB}^3]=[1,2,1]\in\mathbb Z_{15}^3$, and $DB^1$ is the desired file, i.e. $d=1$. 
  \begin{itemize}
  \item {\bf Set up}.
  Let $C_1$ be a cyclic code in  $\mathbb{Z}_3[x]/\langle x^{13}-1\rangle$  and $C_2$ be a cyclic code in  $\mathbb{Z}_5[x]/\langle x^{13}-1\rangle$   with generator polynomial $g_1(x)=x^7+x^5+x^4+2x^3+2x^2+2$ and  $g_2(x)=x^9+2x^8+4x^7+3x^5+2x^4+x^2+3x+4$, respectively.  One can check that $C_\mathtt{IN}=CRT(C_1,C_2)$ is a cyclic code in  $\mathcal R=\mathbb{Z}_{15}[x]/\langle x^{13}-1\rangle$ with generator polynomial $g_\mathtt{IN}=6x^9+12x^8+4x^7+13x^5+7x^4+5x^3+2x^2+3x+14 $. 
  
  Let $\Tilde{C}_1$ be a cyclic code in  $\mathcal R$ with generator polynomial $\tilde{g_1}=10x^7 + 5x^5 + 6x^4 + x^3 + 14x^2 + x + 11$ and let $\Tilde{C}_2$ be a cyclic code  with generator polynomial $\Tilde{g_2}=10x^7 + 11x^5 + 13x^3 + 2x^2 + 10x + 14$. We have that $\Tilde{C}_2\subseteq \Tilde{C}_{1}$. Finally, we select the 2-generator quasi-cyclic code $C_\mathtt{OUT}=[\Tilde{C}_{1},\Tilde{C}_{2}]  \left(\begin{smallmatrix}
  1 & 1  \\
0 & 1
\end{smallmatrix}\right)$ with  generator matrix over  $\mathcal R$ given by
$$G_\mathtt{OUT}=\left(\begin{matrix}
  10x^7 + 5x^5 + 6x^4 + x^3 + 14x^2 + x + 11, & 10x^7 + 5x^5 + 6x^4 + x^3 + 14x^2 + x + 11  \\
0 & 10x^7 + 11x^5 + 13x^3 + 2x^2 + 10x + 14 
\end{matrix}\right).$$ 

\item {\bf Query generation}. Now, consider random polynomials that provide the user the entries of $\mathbf A$ in $\mathcal R=\mathbb{Z}_{15}[x]/\langle x^{13}-1\rangle$, the elements  $e^1_{kj}(x) \in C_\mathtt{IN}$ for $k,j\in\{1,2\}$, and $u^1_{11}(x)$ in $\Tilde{C}_{2} \cap (C_\mathtt{IN}^{\bot}\setminus C_\mathtt{IN}) $  given by \\

    \begin{center}
    \begin{tabular}{rl}\hline
       $a^1_{11}(x)=$  & $5x^{12} + 11x^{11} + 10x^{10} + 13x^9 + 7x^8 +x^7+ 3x^6 + 14x^5 + 14x^4 + 7x^2 + 6x + 4$\\
        $a^1_{12}(x)=$ & $8x^{12} + 8x^{11} + 12x^{10} + 13x^9 + 11x^8 + 11x^7 + 6x^6 + x^5 + 2x^4 + 6x^3 + 5x^2 + 9x + 3$\\
        $a^1_{13}(x)=$ & $5x^{12} + 13x^{11} + 4x^9 + 2x^8 + 14x^7 + 12x^6 + 10x^5 + 6x^3 + 5x^2 + 12x$  \\
        $a^1_{21}(x)=$ & $10x^{12} + 4x^{11} + 4x^9 + 14x^8 + 8x^7 + 6x^5 + 7x^4 + 12x^3 + 11x^2 + 6x + 9$\\
        $a^1_{22}(x)=$ & $5x^{12} + 3x^{11} + 4x^{10} + 10x^9 + 4x^8 + 9x^7 + 14x^6 + 12x^5 + 8x^4 + 9x^3 + 6x^2 + 6x$ \\
        $a^1_{23}(x)=$ & $9x^{12} + 14x^{11} + x^{10} + 4x^9 + x^8 + 13x^7 + 8x^6 + 3x^5 + 13x^4 + 11x^3 + 2x^2 + 14x + 10$\\ \hline
 $e^1_{11}(x)=$  & $11x^{12} + 9x^{11} + 14x^{10} + x^9 + 14x^8 + 12x^7 + 12x^6 + 6x^5 + 12x^4 + 11x^3 + 7x^2 + 2x + 9$\\
 $e^1_{12}(x)=$  & $6x^{12} + 14x^{11} + 10x^{10} + 13x^9 + 3x^8 + 7x^7 + 5x^6 + 3x^5 + 2x^4 + 2x^3 + 10x^2 + 6x + 9$ \\
 $e^1_{13}(x)=$  & $7x^{12} + 6x^{11} + 9x^{10} + 4x^9 + 13x^8 + 3x^7 + 10x^6 + 2x^5 + 2x^4 + 11x^3 + x^2 + 9x + 13$ \\
 $e^1_{21}(x)=$  & $5x^{12} + 3x^{11} + 13x^{10} + 4x^9 + 9x^8 + 6x^7 + 12x^6 + 13x^5 + 14x^4 + 6x^3 + 6x^2 + 12x + 2$ \\
 $e^1_{22}(x)=$  & $6x^{12} + 14x^{11} + 4x^{10} + 2x^9 + x^8 + x^7 + 7x^6 + 14x^5 + 14x^4 + 11x^3 + 13x^2 + 5x + 13$\\
 $e^1_{23}(x)=$  & $10x^{12} + 5x^{11} + 6x^{10} + 3x^9 + 11x^8 + 9x^7 + 13x^6 + 12x^4 + x^3 + 14x^2 + 7x + 14$ \\
        \hline
     $u^1_{11}(x)=$ &  $13x^{12} + 14x^{11} + x^{10} + 3x^9 + 2x^8 + 2x^7 + 7x^6 + 7x^5 + 13x^4 + 6x^3 + 4x^2 + 3x$ \\
     \hline
    \end{tabular}
     \end{center}

\

Then the user can compute  $\boldsymbol{\Delta}$ given by\\

  \begin{tabular}{rl}\hline
       $\delta^1_{11}(x)=$  & $10x^{12} + 9x^{11} + 10x^{10} + 5x^9 + 9x^8 + 13x^7 + 9x^6 + 9x^5 + 12x^4 + 12x^3 + 11x^2 + 6x + 5 $\\
        $\delta^1_{12}(x)=$ & $    13x^{12} + 7x^{11} + 14x^{10} + 13x^9 + 6x^8 + 3x^7 + 5x^6 + 13x^5 + 7x^4 + x^3 + 12x^2 + 11x $\\
        $\delta^1_{13}(x)=$ & $6x^{12} + 6x^{11} + 12x^{10} + 8x^9 + 2x^8 + 10x^7 + 13x^6 + 13x^5 + 10x^4 + 10x^3 + 6x^2 + 9x + 9$  \\
        $\delta^1_{21}(x)=$ & $9x^{12} + 4x^{11} + 14x^{10} + 2x^9 + 4x^8 + 4x^7 + 6x^6 + 4x^5 + x^3 + 3x^2 + 7x + 11$\\
        $\delta^1_{22}(x)=$ & $4x^{12} + 9x^{11} + 12x^{10} + 8x^9 + 10x^8 + 14x^7 + 5x^6 + 7x^5 + 11x^4 + 3x^2 + 8x + 14 $ \\
        $\delta^1_{23}(x)=$ & $6x^{12} + 10x^{11} + 7x^{10} + 8x^9 + 11x^7 + 11x^6 + 13x^5 + 11x^4 + 6x^3 + 5x + 2$\\ 
\hline
    \end{tabular}
\\[1em]
and   the query matrix $\mathbf{Q}$ is the result of expanding in $\mathbb Z_{15}$ the entries in $(\mathbf{A}\mathbin\Vert \boldsymbol{\Delta})$

$$\tiny \left[\begin{array}{l} 4,6,7,0,14,14,3,1,7,13,10,11,5,3,9,5,6,2,1,6,11,11,13,12,8,8,5,6,11,12,12,9,9,13,9,5,10,9,10,0,11,12,1,7,13,5,3,6,13,14,7,13\\
0,12,5,6,0,10,12,14,2,4,0,13,5,9,6,11,12,7,6,0,8,14,4,0,4,10,9,9,6,10,10,13,13,10,2,8,12,6,6,11,7,3,1,0,4,6,4,4,2,14,4,9\\
0,6,6,9,8,12,14,9,4,10,4,3,5,10,14,2,11,13,3,8,13,1,4,1,14,9,14,8,3,0,11,7,5,14,10,8,12,9,4,2,5,0,6,11,13,11,11,0,8,7,10,6\\
 \end{array}\right]. $$

\item {\bf Server response}.
The server computes $\mathbf{R}=\mathbf{DB}\cdot \mathbf{Q}$, which is equal to 
$$\tiny \left[\begin{array}{c} 4 , 6 , 8 , 6 , 7 , 1 ,11 , 8 , 0 , 1 ,14 ,10 , 5 , 1 , 5, 14, 11, 14 , 1 ,14 ,10, 10, 10 ,13,  0 , 7, 7,  2, 11 , 2, 13, 12 ,10 , 2 , 8 ,14,  1 , 0 ,11 , 9,  0,  3,  9 , 3  ,4 ,13 , 7 ,14 ,10 , 4 ,10 , 7\end{array}\right].$$

\item {\bf Recovering stage}. Once received $\mathbf{R}$, the user can extract the submatrices  $\mathbf{R_1}=(r^1_{1,1},r^1_{1,2})$ and $\mathbf{R_2}=(r^2_{1,1},r^2_{1,2})$ with elements in $\mathcal R=\mathbb{Z}_{15}[x]/\langle x^{13}-1\rangle$ given by the following elements 
\\
\begin{center}
    \begin{tabular}{c|l}\hline
    $\mathbf{R_1}$  &    $ r^1_{1,1}=5x^{12} + 10x^{11} + 14x^{10 }+ x^9 + 8x^7 + 11x^6 + x^5 + 7x^4 + 6x^3 + 8x^2 + 6x + 4$\\
       &  $r^1_{1,2} =7x^{12} + 13x^{10} + 10x^9 + 10x^8 + 10x^7 + 14x^6 + x^5 + 14x^4 + 11x^3 + 14x^2 + 5x + 1$\\ \hline
     $\mathbf{R_2}$     & $r^2_{1,1}=11x^{12} + x^{10} + 14x^9 + 8x^8 + 2x^7 + 10x^6 + 12x^5 + 13x^4 + 2x^3 + 11x^2 + 2x + 7 $  \\
       & $r^2_{1,2}=7x^{12} + 10x^{11} + 4x^{10} + 10x^9 + 14x^8 + 7x^7 + 13x^6 + 4x^5 + 3x^4 + 9x^3 + 3x^2 + 9$      \\ \hline
        \end{tabular}
        \end{center}   
        \vspace{1em}
        
and compute  $\mathbf{R_2}-(\mathbf{R_1}\cdot G_\mathtt{OUT})$,  which is a $1\times 2$ matrix whose elements are 
\\
\begin{center}
    \begin{tabular}{l}\hline
        $14x^{12} + 4x^{11} + 7x^{10} + 14x^9 + 13x^8 + 6x^7 + x^6 + x^5 + 13x^4 + 5x^3 + 11x^2 + 13x + 3 $\\
        $ 11x^{12} + 10x^{11} + 12x^{10} + 9x^9 + 2x^8 + 13x^7 + 12x^6 + 14x^5 + 12x^4 + 6x^2 + 7x + 12$ \\ \hline
        \end{tabular}
        \end{center}

\

The user has selected $\delta=1$ as a column position and the files have one column, hence the user does not need the second position. Thus, the user multiplies each entry of the matrix (written as a vector in $\mathbb Z_{15}^{13}$) by $H^{\top}_{\mathtt{IN} }$, the parity check matrix of the code $C_\mathtt{IN}$, given by

$$H^{\top}_{\mathtt{IN} } = \footnotesize \left[\begin{array}{ccccccccccccc}  1&0&0&0&0&0&0&1&0&2&11&2&0\\
0&1&0&0&0&0&0&0&1&0&2&11&2\\
0&0&1&0&0&0&0&1&0&0&2&4&2\\
0&0&0&1&0&0&0&1&1&14&2&4&10\\
0&0&0&0&1&0&0&2&1&8&0&6&13\\
0&0&0&0&0&1&0&2&2&5&0&4&6\\
0&0&0&0&0&0&1&0&2&11&2&0&1\\
0&0&0&0&0&0&0&3&0&3&9&6&6\\
0&0&0&0&0&0&0&0&3&9&0&9&3
 \end{array}\right]^T. 
$$Therefore, the user computes
$$[3,13,11,5,13,1,1,6,13,14,7,4,14]\cdot H^{\top}_{\mathtt{IN}}= [2,7,0,0,11,14,14,6,3]  $$
Since the user knows $\mathbf{DB}^1\cdot \mathbf{U}\cdot H^{\top}_{\mathtt{IN} }$ and $\mathbf{U}\cdot H^{\top}_{\mathtt{IN} }$, the user can solve the two systems of equations, originally over $\mathbb{Z}_{15}$, that arise when solving the system of equations $(\mathbf{DB}^1\cdot \mathbf{U}\cdot H^{\top}_{\mathtt{IN} })= x\cdot(\mathbf{U}\cdot H^{\top}_{\mathtt{IN} })$ over  $\mathbb{Z}_3$ and $\mathbb{Z}_5$,  respectively. Then the user lifts both solutions to a single solution in $\mathbb Z_{15}$ via the  Chinese remainder theorem, obtaining the desired file $x=\mathbf{DB}^1=1$.
 \end{itemize}
\end{example}

\section{Analysis}\label{sec:4}
\subsection{ PIR Rate}

We recall that the database is an $L \times tr$ matrix whose elements are in $\mathbb{Z}_{m^\prime}$. The entries of the query matrix are elements in the polynomial ring $\mathcal{R}$, but they are sent to the server as elements in $R$ since $\mathcal R$ is an algebra over $R$, therefore the size of the matrix $\mathbf{Q}$ is  $tr \times 2ns $ and its elements are in  $\mathbb{Z}_m$, thus the upload cost is$$H(\mathbf{Q})= 2\cdot t\cdot r\cdot n\cdot s \log(m).$$

On the other hand,
the desired file has $L$ rows and $r$ columns. Therefore, the size of the desired file is $Lr \log(m^\prime)$. The server computes the matrix product of the database and the query matrix, and hence the download cost is $$H({\mathbf R})= 2\cdot L\cdot n\cdot s \log(m).$$
The {\bf PIR rate} is the ratio of the desired file size over the sum of the download and upload costs, in the protocol we propose  is given by
\begin{equation}
    \dfrac{Lr \log(m^\prime) }{2\cdot t\cdot r\cdot n\cdot s \log(m) + 2\cdot L\cdot n\cdot s \log(m)}= \dfrac{L\cdot r}{t\cdot r+L} \left( \dfrac{ \log(m^\prime)}{2\cdot n\cdot s \log(m)}\right).
\end{equation}
As usual, we assume that the size of the files is much larger than the number of files, $L\gg tr$, hence the PIR rate of the protocol is approximately equal to
\begin{equation}
    \hbox{PIR rate }\approx \dfrac{r}{2\cdot n\cdot s}\cdot \dfrac{ \log(m^\prime)}{ \log(m)}.
\end{equation}

\begin{table}[h!]
\centering
\begin{tabular}{c c c c c c c} 
 \hline
 $m$ & $n$ & $s$ & $r$ & $k$ & $R$ & $S$ \\ [0.5ex] 
 \hline\hline
 36 & 91 & 5 & 4 & 377 & $\frac{1}{455}$ & $\geq (2^{28})^{6}$ \\ [0.5ex] 
 36 & 91 & 5 & 5 & 377 & $\frac{1}{364}$ & $\geq (2^{28})^{6}$ \\
 36 & 91 & 6 & 6 & 435 & $\frac{1}{364}$ &  $\geq (2^{28})^{7}$\\
 36 & 91 & 10 & 10 & 607 & $\frac{1}{364}$ & $\geq (2^{28})^{11}$ \\
 216 & 91 & 5 & 5 & 377 & $\frac{1}{546}$ &  $\geq ( 2^{28})^{6}$\\ 
\end{tabular}
\caption{Rate and work factor based on parameter selection. The column labeled 'R' shows rates corresponding to the respective parameters, while in the 'S' column, the inverse probability is displayed as the work factor.}\label{table:1}
\end{table}

\subsection{Computational cost}

 The cost is dominated by the multiplication of matrices $\mathbf{DB}$ and ${\mathbf{Q}}$, both of them with entries in the ring $R=\mathbb Z_m$, that is, the product is performed with modular arithmetic. In the HHWZ PIR protocol \cite{9174138}, they perform that multiplication in large finite fields as  degree-$s$ polynomials (where $s$ is the degree of the extension of the chosen field over the base field). Thus, their multiplication complexity can be expressed as $\log(q)\cdot s\log(q)$. That is, they consider this multiplication as the product over $\mathbb F_{q^{\sqrt{s}}}$ in  \cite{9174138}. If we fix the number of rows $L$ and the number of files, there are $\delta n$  of such field multiplications. In our protocol, the usual multiplication in $\mathbb{Z}_m$  is performed  $2rns$ times. When $\delta$   in \cite{9174138} equals our defined value $r$, which corresponds to the number of columns in the files, and the code length is the same, we perform $2s$ times more multiplications than  \cite{9174138}. However, the multiplications in our protocol are simpler since we use modular arithmetic and \cite{9174138} uses field arithmetic.

During the data recovery process, the procedures employed involve solving linear systems of equations and utilizing the Chinese remainder theorem (CRT). Thus, the complexity of the required operations to recover the desired file is the complexity of solving linear equations and performing CRT.

\subsection{Security Analysis} \label{sec:Security}

The subspace attack in \cite[Section V.A]{9174138} could be also applied as a submodule attack in our case, but it has at least the same complexity as in \cite{9174138}, thus if we chose the same security parameter on the size of the module, such attack is still unfeasible.
In \cite{Bordage},
the authors showed that the HHWZ PIR protocol is not private, since the server can recover in polynomial time the index of the desired file with high probability. The central concept of the attack is that by eliminating rows from the query matrix that correspond to the desired file, it yields a large decrease in the dimension over the vector space spanned by the rows of this punctured matrix. The dimension loss only shows a low (almost negligible) probability when the rows unrelated to the requested file are deleted.
Thus, the attack method relies on comparing the ranks of the query matrix after deleting different rows. Therefore, it is essential to construct the query matrix in such a way that deleting a row does not reveal information about the desired data. We will show that this is the case in our protocol.

\begin{theorem}\label{teo:1}
    Let $C_{\mathtt{OUT}}$ be a linear code in $\mathcal{R}^s$, and $C_{\mathtt{IN}} \subseteq \mathcal{R} $ an $R$-linear code.  Let  $\boldsymbol{\Delta}$ be chosen as in Section~\ref{sec: 3} and $\boldsymbol{\Delta}={\mathbf W} + {\mathbf E} + {\mathbf U}$ the decomposition in Equation~(\ref{delta}). Then we have that  \begin{equation}\mathrm{rowspan([\boldsymbol{\Delta}])}\subseteq C_{\mathtt{OUT}}+\Gamma_s(C_{\mathtt{IN}}),\end{equation} where $\mathrm{rowspan}$ is taken in $R$ and $ \Gamma_s(C_{\mathtt{IN}})=[C_{\mathtt{IN}},\ldots,C_{\mathtt{IN}}]\mathrm{Id}_{s}$.
\end{theorem}
\begin{proof}
    Let  $ {[\mathbf W]}$, ${[\mathbf E]}$ and ${[\mathbf U]}$ be the matrices over $R$ containing the expansion of each element in $\mathcal R$ to a tuple in $R$ of the matrices $ {\mathbf W}$, ${\mathbf E}$ and ${\mathbf U}$ in  Equation~(\ref{delta}). Note that the entries of ${[\mathbf E]}$ are random elements in $C_\mathtt{IN}$.  From our technical conditions, we have that $\Tilde{C_j} \cap (C_\mathtt{IN}^{\bot} \setminus C_\mathtt{IN}) \neq \{ {\mathbf 0} \}$, for $j\in \{1,\ldots,s\}$, and the entries of $[{\mathbf U}]$ are elements in $\Tilde{C_s} \cap (C_{\mathtt{IN}}^{\bot} \setminus C_{\mathtt{IN}})$. Hence, the components of    $\boldsymbol{\Delta}={\mathbf W} + {\mathbf E} + {\mathbf U}$ fulfill
\begin{equation*}
\begin{split}
    &\mathrm{rowspan([{\mathbf W}])}  \subseteq C_{\mathtt{OUT}},\\
    &\mathrm{rowspan([{\mathbf E}])}  \subseteq \Gamma_s(C_{\mathtt{IN}}),\\
   &\mathrm{rowspan}([{\mathbf U}]) \subseteq C_{\mathtt{OUT}} \cap \Gamma_s(C_{\mathtt{IN}}^{\bot}).
\end{split}
\end{equation*}
 Thus we have
\begin{equation*}
    \mathrm{rowspan}([\boldsymbol{\Delta}])= \mathrm{rowspan}( [\mathbf{W+E+U}])\subseteq C_{\mathtt{OUT}}+\Gamma_s(C_{\mathtt{IN}}).
\end{equation*}
\end{proof}

\begin{remark}[Choosing the base ring and the projection codes]\label{rem:nh} Let $\mathbf{Q}$ be the query matrix in our protocol, note that it is a matrix with entries in $\mathbb Z_m$. If we know the factorization of $m= \Pi_{i=1}^{\ell} p_i^{e_i}$, we can consider the matrices $\mathbf{Q}_{p_i^{e_i}}$, for $i=1,\ldots, \ell$, given by reducing each entry of $\mathbf{Q}$ modulo  $\mathbb Z_{p_i^{e_i}}$. For $i=1,\ldots, \ell$, we can consider $\mathbf{Q}_{p_i^{e_i}}$ as a generator matrix of a code over $\mathbb Z_{p_i^{e_i}}$ and $\mathbf{Q}$ can be considered as a generator matrix of a code over $\mathbb Z_m$ which is the CRT code of the previous projection components. Suppose that one can apply linear algebra techniques in at least one of the projection codes, that is, there is a projection  $i_0 \in \{1, \ldots, \ell \}$ such that $\mathbf{Q}_{p_{i_0}^{e_{i_0}}}=[\mathbf{A}_{p_{i_0}^{e_{i_0}}}\mid \boldsymbol{\Delta}_{p_{i_0}^{e_{i_0}}} ]$   generates a free code over $\mathbb Z_{p_{i_0}^{e_{i_0}}}$. In this case, the attack in \cite{Bordage} can be applied successfully. That is, in that case, the positions given by $\mathbf{A}_{p_{i_0}^{e_{i_0}}}$ can be seen as a subset of the information set of the projection code and, provided that the attacker has enough rows, the attack in \cite{Bordage} may show a rank difference when one removes the rows with non-zero components in ${\mathbf U}$. Note that this is the case when $e_{i_0}=1$, since  $\mathbb Z_{p_{i_0}^{e_{i_0}}}$ is a field and then the projection code is free.  \\
If $e_{i}>1$, for each $i=1,\ldots, \ell$, then $\mathbb Z_{p_{i}^{e_{i}}}$ is a chain ring with maximal ideal $\langle p_i\rangle$ and nilpotence index $e_i$. Furthermore, a cyclic code over a chain ring is free if and only if it is a Hensel lift, i.e. it is generated by a monic polynomial in $\mathbb Z_{p_{i}^{e_{i}}}[x]/\langle x^n-1\rangle$ (see \cite[Proposition 4.11]{ChainRings}). Therefore, we must consider non-free projection codes for our protocol that we will call non-Hensel lifts. They are provided by choosing the entries of $\mathbf{A}_{p_{i}^{e_{i}}}$ in  $ p_i \mathbb Z_{p_{i}^{e_{i}}}$,  for each  $i=1,\ldots, \ell$, and cyclic codes whose generator set in standard form  (see \cite{ChainRings} for a definition) involves at least one monic polynomial multiplied by a non-zero power of $p$. Thus, we are ensured that the cyclic part is non-free by \cite[Theorem 4.5]{ChainRings}.
\end{remark}
\begin{corollary} There is no $R$-rank difference in the sub-matrix of $\mathbf{Q}$ that we obtain when removing a row of the original matrix if the projection codes are non-Hensel lifts. Therefore, the attack strategy relying on rank difference in \cite{Bordage} does not apply to our proposed protocol.
\end{corollary}
\begin{proof}
  The result follows from Theorem~\ref{teo:1} and Remark~\ref{rem:nh}.  
\end{proof}

As pointed out in the introduction,  single-server PIR protocols cannot be information-theoretical secure, thus there is always some information leakage that an attacker could use to infer which the index of the file the user wants to retrieve. In our protocol, the inner code, $C_\mathtt{IN}$, and  the codes $ \Tilde{C_i} $, for $i=1,\ldots, s$, are kept as secret information. The query matrix $\mathbf{Q}$ is public information, meaning the attacker knows $[\boldsymbol{\Delta}]$ the $\mathbb Z_m$-expansion of  $\boldsymbol{\Delta}$.  Therefore,  the attacker may obtain some information on the constituent cyclic codes by considering their possible defining $f$-cyclotomic cosets where $f$ is a factor of the length of  $[\boldsymbol{\Delta}]$. However, it is still difficult for an attacker to find the inner and outer codes using brute force attacks. Indeed, since they are non-free codes, their generator matrix has linearly dependent vectors. Hence, there is no information set for these codes. Thus,  straight linear algebra operations can not be applied to find these codes.

\begin{remark}
   The cyclotomic coset containing $\theta \in \mathbb{Z}_n$, denoted by $U_\theta$, is the set $\{\theta, \theta q,\dots, \theta q^i\}\pmod{n}$ where $i$ is the smallest integer such that $q^i\equiv 1 \pmod{n}$. It is well known that the number of irreducible factors of $x^n-1$ over $\mathbb{F}_q$ is equal to the number of cyclotomic cosets of $q$ modulo $n$ and each of those factors is called a cyclotomic polynomial. They are
    $
        \Phi_n(x) = \prod_{d \mid n}(x^d -1)^{\mu(\frac{n}{d})}
    $, where where $\mu$ is the M\"obius function. 
 
    Let $(q,n)=1$  and let  $\varphi(n)$ be the Euler phi function. One has that $\Phi_n(x)$ can be factorized into $\varphi(n)/m$ distinct monic irreducible polynomials of degree $m$ over $\mathbb{F}_q$, where $m$ is the least positive integer such that $q^m \equiv 1 ~(\mod{n})$ and $m$ is called the order of $q$ modulo $n$ and it is denoted as $\mathrm{ord}_d(q)$. Therefore, the  number of cyclotomic cosets is
    $
        T=\sum_{d \vert n} \dfrac{\varphi(d)}{\mathrm{ord}_d(q)}.
    $
\end{remark}
 When generating the protocol, we can assume that the user selects $x^n-1$ with a number of irreducible factors big enough. In this way, protection against brute force attacks can be provided. 
From  Theorem \ref{teo:1}, one has that $\mathrm{rowspan}([\boldsymbol{\Delta}]) \subseteq  C_{\mathtt{OUT}}+\Gamma_s(C_{\mathtt{IN}})$. Even though $C_{\mathtt{OUT}}+\Gamma_s(C_{\mathtt{IN}})$ is known, it is hard to get $C_{\mathtt{OUT}}$ and $C_{\mathtt{IN}}$. First, the attacker must use the Chinese remainder theorem to decompose the code $C_{\mathtt{OUT}}+\Gamma_s(C_{\mathtt{IN}})$ into a direct sum of ideals over $\mathbb{Z}_{p_i^{e_i}}[x]/\langle x^n -1\rangle$. Then the attacker must compute the number of cyclotomic cosets in order to find the number of cyclic codes for each decomposition. For instance, if $T$ is the number of cyclotomic cosets, then there are $2^T$ divisors of $x^n-1$, equal to the number of cyclic codes of length $n$. As we mentioned, the attacker could get some cyclotomic polynomials from $\mathrm{rowspan}([\boldsymbol{\Delta}])$. Suppose we define $\Tilde{T}$ as the number of cyclotomic polynomials that an attacker cannot obtain without a brute force attack and that there exists at least $2^{ \Tilde{T}}$ cyclic codes of length $n$. In that case, the probability of recovering the code is given by $Pr \leq  \dfrac{1}{2^{\Tilde{T}}}$. Note that we are bounding the total number of cyclic codes by the number of cyclotomic classes since the generating set in standard form of a cyclic code of length $n$ over $\mathbb Z_{p^s}$ is of the form $\{ f_0, pf_1,\ldots, p^{s-1}f_{s-1}\}$, where $f_{s-1}|\ldots|f_1|f_0|x^n-1$ and they are monic polynomials (see \cite[Definition 4.1]{ChainRings}). In other words, the probability of finding the code is inversely proportional to the number of cyclotomic polynomials the attacker cannot obtain. Therefore, that probability  {exponentially} decreases as $\Tilde{T}$  increases. For example, considering the data presented in Table \ref{table:1}, wherein the inner code length $n=91$ over the ring  $\mathbb{Z}_4$, the number of cyclic codes is at least  $2^{10}$, and over $\mathbb{Z}_{9}$, the number of cyclic codes is at least $2^{18}$. .
For a specific choice of component $s=10$, the probability that an attacker could successfully guess the outer codes is determined by $\mathrm{Pr}\leq (2^{-28})^{10}$ and the probability of guessing the inner codes is determined by $\mathrm{Pr}\leq  2^{-28}$. 
Therefore, the probability of guessing all codes is $\mathrm{Pr}\leq (2^{-28})^{11}$. This probability is notably low, underscoring the resilience of the system against brute force attacks.\\


%


\section*{Conclusions}
In this paper, we introduce a Private Information Retrieval (PIR) protocol designed for a single server, offering computational security derived from coding theory. Despite this protocol having a higher rate compared to the one proposed in \cite{9174138}, it addresses the primary security vulnerability highlighted in \cite{Bordage}. Our proposed protocol is broadly applicable from the coding theory standpoint, covering both rings and fields. We particularly develop it in the case where the inner code is defined over the ring $R=\mathbb Z_m$ and the outer code is defined over $\mathcal R=\mathbb Z_m[x]/\langle x^n-1 \rangle$. This choice of rings ensures that the arithmetic carried out by the server and the user remains purely modular (over $\mathbb Z_m$), presenting an advantage over the field-based alternative that demands larger non-prime fields and, consequently, field arithmetic.

\section*{Acknowledgment}

The authors would like to thank Camilla Hollanti, Ragnar Freij-Hollanti and Neehar Verma for useful discussions.

\ifCLASSOPTIONcaptionsoff
  \newpage
\fi



\bibliographystyle{plain}
\bibliography{SingleServerPIR.bib}

\begin{thebibliography}{10}

\bibitem{angel2018pir}
Sebastian Angel, Hao Chen, Kim Laine, and Srinath Setty.
\newblock Pir with compressed queries and amortized query processing.
\newblock In {\em 2018 IEEE symposium on security and privacy (SP)}, pages
  962--979. IEEE, 2018.

\bibitem{banawan2018capacity}
Karim Banawan and Sennur Ulukus.
\newblock The capacity of private information retrieval from coded databases.
\newblock {\em IEEE Transactions on Information Theory}, 64(3):1945--1956,
  2018.

\bibitem{2002Breaking}
A.~Beimel, Y.~Ishai, E.~Kushilevitz, and J.~F. Raymond.
\newblock Breaking the o(n1/(2k-1)) barrier for information-theoretic private
  information retrieval.
\newblock In {\em Foundations of Computer Science, 2002. Proceedings. The 43rd
  Annual IEEE Symposium on}, 2002.

\bibitem{Bhowmick2020}
Sanjit Bhowmick, Alexandre Fotue-Tabue, Edgar Mart{\'{\i}}nez-Moro, Ramakrishna
  Bandi, and Satya Bagchi.
\newblock Do non-free {LCD} codes over finite commutative frobenius rings
  exist?
\newblock {\em Designs, Codes and Cryptography}, 88(5):825--840, 2020.

\bibitem{blackmore2001matrix}
Tim Blackmore and Graham H.Norton.
\newblock Matrix-product codes over $\mathcal{F}_q$.
\newblock {\em Appl. Algebra Engrg. Comm. Comput.}, 12:477--500, 2001.

\bibitem{Bordage}
Sarah Bordage and Julien Lavauzelle.
\newblock On the privacy of a code-based single-server computational pir
  scheme.
\newblock {\em Cryptogr. Commun.}, 13:519--526, 2020.

\bibitem{Sudan}
Benny Chor, Eyal Kushilevitz, Oded Goldreich, and Madhu Sudan.
\newblock Private information retrieval.
\newblock {\em J. ACM}, 45(6):965–981, nov 1998.

\bibitem{Dvir}
Zeev Dvir and Sivakanth Gopi.
\newblock 2-server pir with sub-polynomial communication.
\newblock In {\em Proceedings of the Forty-Seventh Annual ACM Symposium on
  Theory of Computing}, STOC '15, page 577–584, New York, NY, USA, 2015.
  Association for Computing Machinery.

\bibitem{freij2017private}
Ragnar Freij-Hollanti, Oliver~W Gnilke, Camilla Hollanti, and David~A Karpuk.
\newblock Private information retrieval from coded databases with colluding
  servers.
\newblock {\em SIAM Journal on Applied Algebra and Geometry}, 1(1):647--664,
  2017.

\bibitem{gentry2019compressible}
Craig Gentry and Shai Halevi.
\newblock Compressible fhe with applications to pir.
\newblock In {\em Theory of Cryptography Conference}, pages 438--464. Springer,
  2019.

\bibitem{9174138}
Lukas Holzbaur, Camilla Hollanti, and Antonia Wachter-Zeh.
\newblock Computational code-based single-server private information retrieval.
\newblock In {\em 2020 IEEE International Symposium on Information Theory
  (ISIT)}, pages 1065--1070, 2020.

\bibitem{kiayias2015optimal}
Aggelos Kiayias, Nikos Leonardos, Helger Lipmaa, Kateryna Pavlyk, and Qiang
  Tang.
\newblock Optimal rate private information retrieval from homomorphic
  encryption.
\newblock {\em Proc. Priv. Enhancing Technol.}, 2015(2):222--243, 2015.

\bibitem{kushilevitz1997replication}
Eyal Kushilevitz and Rafail Ostrovsky.
\newblock Replication is not needed: Single database, computationally-private
  information retrieval.
\newblock In {\em Proceedings 38th annual symposium on foundations of computer
  science}, pages 364--373. IEEE, 1997.

\bibitem{lipmaa2017simpler}
Helger Lipmaa and Kateryna Pavlyk.
\newblock A simpler rate-optimal cpir protocol.
\newblock In {\em International conference on financial cryptography and data
  security}, pages 621--638. Springer, 2017.

\bibitem{affine}
E.~Mart\'{\i}nez-Moro, A.~Pi\~{n}era Nicol\'{a}s, and I.~F. R\'{u}a.
\newblock Codes over affine algebras with a finite commutative chain
  coefficient ring.
\newblock {\em Finite Fields Appl.}, 49:94--107, 2018.

\bibitem{melchor2016xpir}
Carlos~Aguilar Melchor, Joris Barrier, Laurent Fousse, and Marc-Olivier
  Killijian.
\newblock Xpir: Private information retrieval for everyone.
\newblock {\em Proceedings on Privacy Enhancing Technologies}, pages 155--174,
  2016.

\bibitem{ChainRings}
Graham~H. Norton and Ana S\u{a}l\u{a}gean.
\newblock On the structure of linear and cyclic codes over a finite chain ring.
\newblock {\em Appl. Algebra Engrg. Comm. Comput.}, 10(6):489--506, 2000.

\bibitem{sion2007computational}
Radu Sion and Bogdan Carbunar.
\newblock On the computational practicality of private information retrieval.
\newblock In {\em Proceedings of the Network and Distributed Systems Security
  Symposium}, pages 2006--06. Internet Society Geneva, Switzerland, 2007.

\bibitem{sun2017capacity}
Hua Sun and Syed~Ali Jafar.
\newblock The capacity of robust private information retrieval with colluding
  databases.
\newblock {\em IEEE Transactions on Information Theory}, 64(4):2361--2370,
  2017.

\bibitem{sun2018capacity}
Hua Sun and Syed~Ali Jafar.
\newblock The capacity of symmetric private information retrieval.
\newblock {\em IEEE Transactions on Information Theory}, 65(1):322--329, 2018.

\bibitem{chatgpt}
Imdad Ullah, Najm Hassan, Sukhpal~Singh Gill, Basem Suleiman, Tariq~Ahamed
  Ahanger, Zawar Shah, Junaid Qadir, and Salil~S. Kanhere.
\newblock Privacy preserving large language models: Chatgpt case study based
  vision and framework.
\newblock arXiv.2310.12523, 2023.

\bibitem{asch}
Bram van Asch.
\newblock Matrix-product codes over finite chain rings.
\newblock {\em Appl. Algebra Engrg. Comm. Comput.}, 19(1):39--49, 2008.

\bibitem{yi2012single}
Xun Yi, Mohammed~Golam Kaosar, Russell Paulet, and Elisa Bertino.
\newblock Single-database private information retrieval from fully homomorphic
  encryption.
\newblock {\em IEEE Transactions on Knowledge and Data Engineering},
  25(5):1125--1134, 2012.

\end{thebibliography}
\end{document}